\documentclass[12pt,a4paper]{article}
\pdfoutput=1
\usepackage{jheppub}
\usepackage{hyperref, amsmath, amssymb, graphicx, amsfonts, latexsym, bbold, mathbbol}
\usepackage{tabstackengine}
\usepackage{color}
\newcommand{    \be}{\begin{equation}}
\newcommand{\ee}{\end{equation}}
\newcommand{\bea}{\begin{eqnarray}}
\newcommand{\eea}{\end{eqnarray}}
\def\eqa{&=&} 
\def\ccr{\nonumber\\} 
\newcommand{\dslash}{\partial \! \! \!  /} 
\newcommand{\DDslash}{D \! \! \! \!  / \, } 
\newcommand{\bpsi}{\overline{\psi}}
\newcommand{\blambda}{\overline{\lambda}}
\newcommand{\brho}{\overline{\rho}}

\def\la{\langle}
\def\ra{\rangle}
\def\tr{{\rm tr}\,}
\author[a,b,c]{Fiorenzo Bastianelli,}
\author[a,c]{Matteo Broccoli,}

\affiliation[a] {Dipartimento di Fisica ed Astronomia, Universit{\`a} di Bologna,
via Irnerio 46, I-40126 Bologna, Italy}
\affiliation[b] {INFN, Sezione di Bologna, via Irnerio 46, I-40126 Bologna, Italy}
\affiliation[c]{Max-Planck-Institut f\"ur Gravitationsphysik (Albert-Einstein-Institut)\\
 Am M\"uhlenberg 1, D-14476 Golm, Germany}
 \emailAdd{bastianelli@bo.infn.it} 
 \emailAdd{matteo.broccoli2@studio.unibo.it}
 
\abstract{We study the trace anomaly of a Weyl fermion in an abelian gauge background. 
Although the presence of the chiral anomaly implies a breakdown of gauge invariance, we find that the trace anomaly 
can be cast in a gauge invariant form. In particular, we find that it does not contain any  odd-parity
contribution proportional to the Chern-Pontryagin density, which would be allowed by the consistency conditions.
We perform our calculations using Pauli-Villars regularization and heat kernel methods.
The issue is analogous to the one recently discussed in the literature about the trace anomaly 
of a Weyl fermion in curved backgrounds.
}

\title{On the trace anomaly of a Weyl fermion in a gauge background}
\begin{document}
\maketitle
\flushbottom


\section{Introduction}
In this paper we study the trace anomaly of a chiral fermion coupled  to an abelian
gauge field in four dimensions. It is  well-known that the model contains  an anomaly in the axial gauge  symmetry,
thus preventing the quantization of  the gauge field in a consistent manner.  Nevertheless,
it is useful to study the explicit structure of the trace anomaly emerging in the axial $U(1)$ background.

One reason to study the problem is that an analogous situation has recently been addressed for a Weyl fermion coupled to gravity. 
In particular, the presence of an odd-parity term in the trace anomaly (the Pontryagin density of the curved background) 
has been reported in \cite{Bonora:2014qla}, and further elaborated upon in \cite{Bonora:2017gzz, Bonora:2018obr}.
This anomaly was also envisaged in \cite{Nakayama:2012gu},  and discussed more recently in \cite{Nakayama:2018dig}. 
However, there are many indications  that such an anomaly cannot be present in the theory of a Weyl fermion. 
The explicit calculation  carried out in \cite{Bastianelli:2016nuf} confirms this last point of view. 

One of the reasons why one does not expect the odd-parity contribution to the trace anomaly is that by CPT in four dimensions a
left handed fermion has a right handed antiparticle, expected to contribute oppositely to any chiral imbalance in the coupling 
to gravity. To see that, one may cast the quantum field theory of a Weyl fermion as the quantum theory of a Majorana fermion.
The latter shows no sign of an odd-parity trace anomaly. Indeed, the functional determinant that arises in a path integral quantization can be 
regulated using Pauli-Villars Majorana fermions with Majorana mass, so to keep the determinant manifestly real in euclidean space, 
thereby excluding the appearance of any phase that might 
produce an anomaly (the odd-parity term carries an imaginary coefficient in euclidean space) \cite{AlvarezGaume:1983ig}.
Recently,  this has been  verified again using Feynman diagrams \cite{Godazgar:2018boc}, confirming the results of   \cite{Bastianelli:2016nuf}.
 An additional piece of evidence comes from studies of the 3-point functions of conserved currents in four dimensional
 CFT, which exclude odd-parity terms in the correlation function of three stress tensors  
 at non-coinciding points \cite{Stanev:2012nq,Zhiboedov:2012bm}, 
 seemingly excluding its presence also in the trace anomaly (see however \cite{Bonora:2015nqa}).
 
Here we analyze the analogous situation of a chiral fermion coupled to an abelian $U(1)$ gauge background. As well-known
the theory exhibits a chiral anomaly, that implies a breakdown of gauge invariance. 
It is nevertheless interesting to compute its trace anomaly as well.
Apart from the standard gauge invariant contribution ($\sim F^2$) and  possible gauge noninvariant terms, which as we shall show 
can be canceled by counterterms, one might expect a contribution from the odd-parity Chern-Pontryagin density $F\tilde F$.  Indeed  the latter 
satisfies the consistency conditions for trace anomalies. In addition, the fermionic functional determinant is complex in euclidean space,
and thus carries a phase (responsible for the $U(1)$ axial anomaly). On the other hand,  the structure of the 3-point function 
of the stress tensor with two $U(1)$ currents in generic CFTs does not allow for odd-parity terms \cite{Stanev:2012nq,Zhiboedov:2012bm}
that could signal a corresponding anomaly in the trace of the stress tensor in a $U(1)$ background. Thus,
apart from a few differences, the case seems analogous to that of the chiral fermion in curved space, 
and is worth addressing.

To ascertain the situation we compute explicitly the trace anomaly of a Weyl  fermion coupled to a $U(1)$ gauge field.
Using a Pauli-Villars regularization we find that no odd-parity term emerges in the quantum trace of the stress tensor.
We use a Majorana mass for computing the trace anomaly, as this mass term can be covariantized (in curved space)
without the need of introducing additional fields of opposite chirality, which on the other hand would be required by a Dirac mass.
The coupling to gravity through the vierbein (needed only at the linear order) 
is used to treat the vierbein  as an external source for the stress tensor,
and to relate the trace of the latter to a Weyl rescaling of the former. 
The manifest covariance of the Majorana mass guarantees that the stress tensor can be kept 
conserved\footnote{Up to a contribution from the background 
$U(1)$ gauge field, already present at the classical level.} and symmetric 
also at the quantum level, i.e. without general coordinate (Einstein)  and local Lorentz anomalies.
We repeat part of our calculations with a  Dirac mass as well. In addition, we calculate also the anomalies 
of a massless Dirac fermion which, though well-known, serve for comparison and as a test on the scheme adopted.
The final result is that the trace anomaly of a Weyl fermion 
does not contain any odd-parity contribution proportional to the Chern-Pontryagin density, and 
it can be written in a gauge invariant form that is equal to half the trace anomaly of a Dirac fermion.
We verify the consistency of the different regularizations used, and report the  local counterterms that relate them.

We organize the paper as follows. In section \ref{s2} we set up the stage and 
review the lagrangians of the Weyl and  Dirac fermions, respectively,  and identify the relevant 
differential operators that enter our regularization scheme.
In section \ref{s3} we review the method that we choose for computing the chiral and trace
 anomalies. In section \ref{s4}  we present our final results.
We conclude in section \ref{s5}, confining to the appendices notational conventions, heat kernels formulas, and sample calculations.

\section{Actions and symmetries}
\label{s2}

We first present the classical models and review  their main properties to set up the stage for our calculations. 
The model of main interest is a massless  Weyl fermion coupled to an abelian gauge field.
We first describe its symmetries, and then the mass terms to be used in a Pauli-Villars regularization.
For comparison,  we consider also a massless Dirac fermion coupled to vector and axial abelian gauge fields, 
a set-up used by Bardeen to compute systematically the anomalies in vector and axial currents \cite{Bardeen:1969md}. 
Our notation is commented upon and recapitulated in appendix \ref{appA}. 

\subsection{The Weyl fermion}

The lagrangian of a left handed Weyl spinor $\lambda$ coupled to a $U(1)$  gauge field is 
 \be
{\cal L}_{\scriptscriptstyle W} = - \blambda \gamma^a (\partial_a -iA_a) \lambda =
-  \blambda \gamma^a D_a (A) \lambda   = -  \blambda \DDslash (A) \lambda  
\label{Weyl-act}
\ee
where the chirality of the spinor is defined by the constraint $\gamma^5 \lambda =\lambda$, or equivalently $\lambda=\frac{1+\gamma^5}{2}\lambda$.
It is classically gauge invariant and conformally invariant. Both symmetries become anomalous at the quantum level.

In  the following we find it convenient to use the charge conjugated spinor $\lambda_c$, which has the opposite chirality of $\lambda$
\be
\lambda_c = C^{-1} \blambda^T  \;, \qquad \gamma^5\lambda_c = -\lambda_c \;.
\ee
The lagrangian can be cast in equivalent forms using $\lambda_c$ rather then $\blambda$ 
\be
{\cal L}_{\scriptscriptstyle W} =\lambda_c^T C \DDslash (A) \lambda  =  \lambda^T C \DDslash (-A) \lambda_c  = 
\frac12 \left ( \lambda_c^T C \DDslash (A) \lambda +   \lambda^T C \DDslash (-A) \lambda_c \right) 
\ee
with the last two forms valid up to boundary terms (we perform partial integrations in the action and drop boundary terms).
We use the last form in our calculations.

The gauge transformations are 
\be 
	\left\{
	\begin{aligned}
   \lambda(x) \quad &\to\quad \lambda'(x) = e^{i\alpha(x)} \lambda(x)  \\
    \blambda(x) \quad &\to \quad\blambda'(x) = e^{-i\alpha(x)} \blambda(x) \\
    \lambda_c(x) \quad &\to\quad \lambda_c'(x) = e^{-i\alpha(x)} \lambda_c(x)  \\
      A_a(x) \quad &\to \quad  A'_a(x)  =  A_a(x) + \partial_a \alpha (x) 	
		\end{aligned}
	\right.
	\label{axial-transf}
\ee
and the action $S_{\scriptscriptstyle W}=\int d^4 x\, {\cal L}_{\scriptscriptstyle W}$ is gauge invariant. 
Recall also that $A_a$ can be used as an  external source for  the current 
\be
J^a = i\blambda \gamma^a \lambda \;. 
\ee
Varying only $A_a$ in the action with a gauge transformation of infinitesimal parameter $\alpha(x)$ produces
\be
\delta_\alpha^{(A)} S_{\scriptscriptstyle W} = - \int d^4 x \, \alpha(x)  \partial_a J^a(x)
\ee
and the full gauge symmetry ($\delta_\alpha S_{\scriptscriptstyle W} =0$)
guarantees that  the $U(1)$ current is conserved  on-shell (i.e. using the fermion equations of motion)
 \be
 \partial_a J^a(x)=0\;.
 \ee

Similarly, one may check that the action is classically conformal invariant and that the stress tensor has a vanishing trace. To see this,
one couples the model to gravity by introducing the vierbein $e_\mu{}^a$ (and related spin connection $\omega_{\mu}{}^{ab}$), 
and realizes that the action is invariant under general coordinate, local Lorentz, and Weyl  transformations.
The energy momentum tensor, or stress tensor,  is defined by
\be
T^{\mu a}(x) = \frac{1}{e}\frac{\delta S_{\scriptscriptstyle W}}{\delta e_{\mu a} (x)}
\ee
where $e$ is the determinant of the vierbein, 
and is covariantly conserved\footnote{For conservation one needs to use also the 
equations of motion of the gauge field, or alternatively keep the expected gauge field
contribution on the right hand side of the conservation equation, see previous footnote.},
symmetric, and traceless on-shell, as consequence of diffeomorphisms, local Lorentz invariance, and Weyl symmetry,
 respectively
\be
\nabla_\mu T^{\mu a} =0 \;, \qquad T_{ab} =  T_{ba} \;, \qquad   T^a{}_a =0 
\label{cons-1}
\ee
(indices are made ``curved" or ``flat" by using the vierbein and its inverse). The vierbein can be used as an external source for 
 the stress tensor, and an infinitesimal Weyl transformation on the vierbein acts as a  source for the trace $T^a{}_a$ of the stress tensor.
In  the following we only need a linearized coupling to gravity to produce a single insertion of the stress tensor in correlation functions.
Apart from that, we are only interested in flat space results. In any case, the full coupling to gravity reads 
\be
{\cal L}_{\scriptscriptstyle W} = - e\,  \blambda \gamma^\mu  \nabla_\mu \lambda 
\ee
where $\gamma^\mu= e^\mu{}_a  \gamma^a $  are the gamma matrices with curved indices,
$e^\mu{}_a$ is the inverse vierbein, and $\nabla_\mu$ is the covariant derivative containing both the  $U(1)$ gauge field $A_\mu$
and spin connection  $\omega_{\mu ab}$
\be
\nabla_\mu = \partial_\mu -iA_\mu + \frac14 \omega_{\mu ab} \gamma^a\gamma^b  
\;.
\ee
The local Weyl symmetry is given by 
\be 
	\left\{
	\begin{aligned}
	\lambda (x)  \quad &\to\quad  \lambda' (x) = e^{-\frac{3}{2}\sigma (x)} \lambda (x)  \\
	\blambda (x)  \quad &\to\quad  \blambda' (x)  = e^{-\frac{3}{2}\sigma (x)} \blambda (x)  \\
        A_a (x)  \quad &\to\quad  A'_a (x)  = A_a (x)  \\
         e_\mu{}^a(x) \quad &\to\quad  {e'}_\mu{}^a(x) = e^{\sigma (x)} e_\mu{}^a(x)
	\end{aligned}
	\right.
	\label{212}
\ee
where $\sigma(x)$ is an arbitrary function. Varying in the action only the vierbein  with an
infinitesimal Weyl transformation produces  the trace of the stress tensor
\be
\delta_\sigma^{(e)} S_{\scriptscriptstyle W} =  \int d^4 x e \, \sigma(x)   T^a{}_a(x) 
\ee
and the full Weyl symmetry ($\delta_\sigma S_{\scriptscriptstyle W} =0$)
guarantees that  the stress tensor is traceless on-shell 
 \be
 T^a{}_a(x) =0 \;. 
 \ee
 For completeness, we record the form of the stress tensor in flat space 
emerging form the previous considerations and simplified by using the
equations of motion
\be
T_{ab} = \frac14 \blambda 
\left (\gamma_a {\stackrel{\leftrightarrow}{D}}_b + \gamma_b {\stackrel{\leftrightarrow}{D}}_a \right)\lambda 
\ee
where ${\stackrel{\leftrightarrow}{D}}_a  = {\stackrel{}{D}}_a  - {\stackrel{\leftarrow}{D}}_a $ 
(in terms of the gauge covariant derivative). Obviously, it is traceless on-shell.

\subsubsection{Mass terms}

To compute the anomalies in the quantum theory we regularize the latter using massive Pauli-Villars (PV) fields, with the anomalies 
coming eventually from the noninvariance of the mass term.  
For the massless Weyl fermion, one can take as  PV field a Weyl fermion of the same chirality with a Majorana mass added. 
The mass term is Lorentz invariant, but breaks the gauge and conformal symmetries. It takes many equivalent forms
\begin{align}
\Delta_{\scriptscriptstyle M}{\cal L}_{\scriptscriptstyle W} &= \frac{M}{2} \left (\lambda^T C \lambda + {\rm h.c.} \right )
 = \frac{M}{2} \left ( \lambda^T C \lambda -\blambda C^{-1} \blambda^T \right ) 
 \ccr
 &
 =
\frac{M}{2} \left ( \lambda^T C \lambda +\lambda_c^T C \lambda_c \right )
\label{M-mass}
\end{align}
where h.c. denotes the hermitian conjugate and $M$ is a real mass parameter.
 Since the charge conjugation matrix $C$ is antisymmetric this term is nonvanishing for 
anticommuting spinors\footnote{In terms of the 2-component left handed Weyl spinor
 $l_\alpha$
this mass terms reads as
\begin{align}
\Delta_{\scriptscriptstyle M}{\cal L}_{\scriptscriptstyle W} &= \frac{M}{2} 
\left (l_\alpha (-i\sigma^2)^{\alpha\beta}l_\beta +l^*_{\dot \alpha} (i\sigma^2)^{\dot \alpha\dot \beta}l^*_{\dot \beta}
 \right ) 
\end{align}
and does not contain any other spinor apart from $l_\alpha$ and its complex conjugate 
$l^*_{\dot \alpha}$.
In the chiral representation of the gamma matrices the 2-component spinor $l_\alpha$ sits inside $\lambda$ as in eq. \eqref{A.12}. }.

Casting the full  massive PV action  ${\cal L}_{\scriptscriptstyle PV}  ={\cal L}_{\scriptscriptstyle W} + \Delta_{\scriptscriptstyle M}{\cal L}_{\scriptscriptstyle W}$ 
 in the following compact form
\be 
{\cal L}_{\scriptscriptstyle PV} =  \frac12 \phi^T T  {\cal O} \phi +\frac12 M \phi^T T \phi  \;,
\label{gf}
\ee 
where $\phi$ is a column vector containing both
$\lambda$ and $\lambda_c$ ($\phi$ is thus a 8 dimensional vector)
\be
\phi = \left( \begin{array}{c} 
\lambda \\ 
\lambda_c 
\end{array} \right)  \;,
\ee
permits the identification of the operators 
\be
T {\cal O} =
\left( \begin{array}{cc}
  0 &  C \DDslash(-A) P_R 
  \\
C \DDslash(A) P_L  &0  
\end{array} \right)  
\;, \qquad 
T  = \left( \begin{array}{cc}  C  P_L&0  \\  0 &  C P_R  \end{array} \right)  
\ee
and 
\be
{\cal O} =
\left( \begin{array}{cc}
0 & \DDslash(-A) P_R   \\
  \DDslash(A)P_L &0 
\end{array} \right)  
\;, \qquad 
{\cal O}^2 =
\left( \begin{array}{cc}
  \DDslash(-A) \DDslash(A) P_L
  &0 \\
0 & \DDslash(A) \DDslash(-A) P_R
\end{array} \right)  \;.
\label{PV-MM-reg}
\ee
The latter will be used in our anomaly calculations. The chiral projectors $P_L$ and $P_R$ 
\be
P_L = \frac{\mathbb{1}+\gamma^5}{2}  \;, \qquad  P_R = \frac{\mathbb{1}-\gamma^5}{2} 
\ee
have been introduced to stress that the matrix $T$ is not invertible
in the full 8 dimensional space on which $\phi$ lives.
An advantage of the Majorana mass term is that it can be constructed without the need of
introducing extra degrees of freedom (as required by a Dirac mass term).
Moreover, it can be covariantized under Einstein (general coordinate) and local Lorentz symmetries.
The covariantization is achieved by multiplying it with the determinant of the vierbein $e$
\be
\Delta_{\scriptscriptstyle M}{\cal L}_{\scriptscriptstyle W} =
\frac{e M}{2} \left ( \lambda^T C \lambda +\lambda_c^T C \lambda_c \right )\;.
\label{M-mass-cov}
\ee

An alternative mass term is the Dirac mass. To use it one must introduce in addition 
an uncoupled right handed PV fermion $\rho$
(satisfying $\rho =P_R \rho$), so that the full massive PV lagrangian reads 
\be 
\tilde{\cal L}_{\scriptscriptstyle PV} = 
-  \blambda \DDslash (A) \lambda  -   \brho\dslash\rho - M (\blambda\rho + \brho\lambda)
\label{dirac-1}
\ee
or, equivalently,
\begin{align}
\tilde{\cal L}_{\scriptscriptstyle PV} &= 
\frac12 \left ( \lambda_c^T C \DDslash (A) \lambda +   \lambda^T C \DDslash (-A) \lambda_c \right) +
\frac12 \left ( \rho_c^T C  \dslash \rho +   \rho^T C  \dslash \rho_c \right) \ccr
&+ \frac{M}{2} (\lambda_c ^TC\rho + \rho^TC\lambda_c + \rho_c^TC\lambda + \lambda^TC\rho_c) \;.
\label{dirac-mass-mix}
\end{align}
Casting this PV lagrangian in the general form \eqref{gf}, where
\be
\phi = \left( \begin{array}{c} 
\lambda \\ 
\lambda_c\\ 
\rho \\ 
\rho_c\\ 
\end{array}\right)
\ee
with each entry a 4 dimensional Dirac spinor (with chiral projectors), allows to identify
\be
\fixTABwidth{T}
T{\cal O} = \parenMatrixstack{ 
0 & C\DDslash(-A)P_R &0 &0\\ 
C\DDslash(A)P_L & 0 &0 &0\\
0& 0& 0& C\dslash P_L \\
0& 0& C\dslash P_R &0 
}
\ee
\be
T =
\left(
	\begin{array}{cccc}
	0& 0 &0 & CP_L \\
	0& 0& CP_R &0 \\
	0 &CP_R &0 &0 \\
	CP_L &0 &0 &0 \\
	\end{array}
\right)
\ee
and 
\be
\fixTABwidth{T}
{\cal O} =  \parenMatrixstack{
0& 0& \dslash P_R & 0\\ 
0& 0& 0 & \dslash P_L\\ 
\DDslash(A)P_L & 0& 0& 0 \\ 
0 & \DDslash(-A)P_R &0 &0 
}
\ee
\be
\fixTABwidth{T}
{\cal O}^2 = \parenMatrixstack{
\dslash \DDslash (A)P_L &0 &0 &0 \\
0 &\dslash\DDslash (-A) P_R &0 &0 \\
0 &0 &\DDslash (A)\dslash P_R &0 \\
0 &0 &0 &\DDslash (-A)\dslash P_L
} \;.
\label{PV-WDM-reg}
\ee
These differential operators appear also in \cite{AlvarezGaume:1984dr}, where definitions for the determinant of a chiral Dirac operator were  studied
with the purpose of addressing chiral anomalies.

A drawback of the Dirac mass term, as regulator of the Weyl theory, is that one cannot covariantize it  while keeping the  auxiliary right 
handed spinor $\rho$ free in the kinetic term (it cannot be coupled to gravity, otherwise it would not regulate properly 
the original chiral  theory).
One can still use the regularization by keeping $\rho$ free in the kinetic term, 
 but as the mass term breaks the Einstein and local Lorentz symmetries explicitly, one would get anomalies 
in the conservation $(\partial_a T^{ab})$ and antisymmetric part  $(T^{[ab]})$ of the stress tensor. Then,  one is forced to study the counterterms 
that remove the anomalies in the
conservation and symmetry of the stress tensor (this can always be done in 4 dimensions \cite{AlvarezGaume:1983ig, Bardeen:1984pm}), 
and check which trace anomaly one is left with at the end.  As this is rather laborious,
we do not use this mass term to calculate the trace  anomaly in the Weyl theory\footnote{A possibility  to simplify the calculation would be to use the axial metric background 
introduced in \cite{Bonora:2017gzz,Bonora:2018obr}, but here we will not follow this direction either.}.

\subsection{The Dirac fermion}

We consider also the more general model of a massless Dirac fermion coupled to vector and axial $U(1)$ gauge fields $A_a$ and $B_a$.
The lagrangian is 
\begin{align}
{\cal L}_{\scriptscriptstyle D} &= -\bpsi \gamma^a ( \partial_a -i  A_a -i B_a \gamma^5) \psi  = - \bpsi \DDslash(A,B)\psi \ccr
&= \frac12 \psi_c^T C \DDslash(A,B)\psi  + \frac12  \psi^T C \DDslash(-A,B)\psi_c
\label{D-lag}
\end{align}
where the last form is valid up to boundary terms. A chiral projector emerges when $A_a=\pm B_a$, and we use 
this model to address again the issue of the chiral fermion in flat space 
(the limit $A_a= B_a\to \frac{A_a}{2}$ reproduces the massless part of \eqref{dirac-1}).

The lagrangian is invariant under the local $U(1)_V$ vector transformations
\be 
	\left\{
	\begin{aligned}
		\psi(x) \ &\to \  \psi^\prime(x) = e^{i\alpha(x)} \psi(x) \\
		\bpsi(x) \ &\to \ \bpsi^\prime(x) = e^{-i\alpha(x)} \bpsi (x)\\
	       \psi_c(x) \ &\to \ \psi_c^\prime(x) = e^{-i\alpha(x)} \psi_c(x) \\
		A_a(x) \ &\to \ A_a^\prime(x) = A_a(x) +\partial_a\alpha(x) \\
		B_a(x) \ &\to \ B_a^\prime(x) = B_a(x) 
	\end{aligned}
	\label{vec-s}
	\right.
\ee
and local $U(1)_A$ axial  transformations
\be 
	\left\{
	\begin{aligned}
		\psi(x) \ &\to \ \psi^\prime(x) = e^{i\beta(x) \gamma^5} \psi(x) \\
		\bpsi(x) \ &\to \ \bpsi^\prime(x) = \bpsi(x)  e^{i\beta(x) \gamma^5}  \\
                 \psi_c(x) \ &\to \ \psi_c^\prime(x) = e^{i\beta(x)\gamma^5} \psi_c(x) \\
		A_a(x) \  &\to \ A_a^\prime(x) = A_a(x)  \\
		B_a(x) \ &\to \ B_a^\prime(x) = B_a(x)  +\partial_a\beta(x) \;.
\label{ax-s}
	\end{aligned}
		\right.
\ee
Again one can use $A_a$ and $B_a$  as sources for $J^a= i \bpsi \gamma^a \psi$ and $J_5^a= i \bpsi \gamma^a \gamma^5\psi$, respectively.
Under infinitesimal variation of these external sources  one finds
\begin{align}
& \delta_\alpha^{(A)} S_{\scriptscriptstyle D} = - \int d^4 x \, \alpha(x)  \partial_a J^a(x)  \ccr 
& \delta_\beta^{(B)} S_{\scriptscriptstyle D} = - \int d^4 x \, \beta(x)  \partial_a J_5^a(x)  
\end{align}
and the classical gauge symmetries imply that $J^a$ and $J_5^a$ are conserved on-shell
\begin{align}
& \partial_a J^a(x)  =0 \ccr 
& \partial_a J_5^a(x)  =0 \;.
\end{align}

A coupling to gravity shows that the stress tensor is traceless because of the Weyl symmetry.
The Weyl transformations rules are as in \eqref{212}, with in addition the rule that $B_a$ is left invariant.
An infinitesimal Weyl variation on the vierbein produces the trace of the stress tensor
\be
\delta_\sigma^{(e)} S_{\scriptscriptstyle D} = \int d^4 x e \, \sigma(x)   T^a{}_a(x) \;.
\ee
and the Weyl symmetry implies that it vanishes on-shell
\be
T^a{}_a(x) =0 \;.
\ee

\subsubsection{Mass terms}

To regulate the one-loop graphs we introduces massive PV fields.  The standard Dirac mass term
\be
\Delta_{\scriptscriptstyle M} {\cal L}_{\scriptscriptstyle D} = - M \bpsi  \psi =   \frac{M}{2} (\psi^T_c C \psi +\psi^T C \psi_c)
\ee
preserves  vector gauge invariance, and casting the PV lagrangian  
\be
{\cal L}_{\scriptscriptstyle PV} = {\cal L}_{\scriptscriptstyle D} + \Delta_{\scriptscriptstyle M}{\cal L}_{\scriptscriptstyle D} 
\ee
in the form \eqref{gf}, now  with  $ \phi  = \left( \begin{array}{c}    \psi \\  \psi_c    \end{array} \right)  $,
allows to recognize the operators 
\be
T {\cal O} =
\left( \begin{array}{cc}
  0 &  C \DDslash(-A,B)    \\   C \DDslash(A,B)  &0  
\end{array} \right)  \;, \qquad 
T  = \left( \begin{array}{cc}    0& C  \\  C &  0    \end{array} \right)  
\ee
and 
\be
{\cal O} =
\left( \begin{array}{cc}
  \DDslash(A,B)   &0  \\  0&  \DDslash(-A,B)  
\end{array} \right)  \;, \qquad 
{\cal O}^2 =
\left( \begin{array}{cc}
  \DDslash(A,B)^2   &0  \\  0 &  \DDslash(-A,B)^2 
\end{array} \right)  \;.
\label{PVD-Dmass}
\ee
This mass terms mixes the two chiral parts $\lambda$ and $\rho$ of the Dirac fermion $\psi =\lambda +\rho$,
 see eqs. \eqref{dirac-1} or \eqref{dirac-mass-mix}
that makes it immediately visible. 
After covariantization to gravity the decoupling of the two chiralities is not easily achievable, 
and  relations between the trace anomaly 
of a Dirac fermion and the trace anomaly of a  Weyl fermion cannot be studied directly by using the Dirac mass in the PV regularization.

Thus, it is useful to consider a Majorana mass as well. It breaks both vector and axial symmetries
\be
\tilde \Delta_{\scriptscriptstyle M}{\cal L}_{\scriptscriptstyle D} = 
\frac{M}{2}(\psi^T C\psi + {\rm h.c.}) = 
\frac{M}{2}(\psi^TC\psi + \psi^T_cC\psi_c)
\ee
and one finds from the alternative PV lagrangian 
\be
\tilde {\cal L}_{\scriptscriptstyle PV} = {\cal L}_{\scriptscriptstyle D} + \tilde \Delta_{\scriptscriptstyle M}{\cal L}_{\scriptscriptstyle D} 
\ee
the operators 
\be
T {\cal O} =
\left( \begin{array}{cc}
  0 &  C \DDslash(-A,B)    \\   C \DDslash(A,B)  &0  
\end{array} \right)  \;, \qquad 
T  = \left( \begin{array}{cc}   C &  0   \\     0& C \end{array} \right)  
\ee
and 
\be
{\cal O} =
\left( \begin{array}{cc}
   0&  \DDslash(-A,B) \\    \DDslash(A,B)   &0
\end{array} \right) 
 \;, \quad 
{\cal O}^2 =
\left( \begin{array}{cc}
 \DDslash(-A,B)\DDslash(A,B)   &0   \\  0 &  \DDslash(A,B)\DDslash(-A,B) 
\end{array} \right)  \;.
\label{PVD-MM}
\ee
Covariantization to gravity does not mix the chiral parts of the Dirac fermion, and a decoupling limit to the chiral theory of a Weyl fermion $\lambda$ is now attainable.

\section{Regulators and consistent anomalies}
\label{s3}

To compute the anomalies we employ a Pauli-Villars regularization  \cite{Pauli:1949zm}.
Following the scheme of refs. \cite{Diaz:1989nx,Hatsuda:1989qy} we cast the calculation in the same form as the one 
obtained by Fujikawa in analyzing the measure of the path integral \cite{Fujikawa:1979ay,Fujikawa:1980vr}.
This set-up makes it easier to use heat kernel formulas  \cite{DeWitt:1965jb, DeWitt:1985bc} to evaluate the anomalies explicitly. 
At the same time, the method guarantees that one obtains consistent anomalies,
i.e. anomalies that satisfy the consistency conditions \cite{Wess:1971yu,Bonora:1983ff}.

Let us review the scheme of  ref. \cite{Diaz:1989nx}. One considers a lagrangian for a field $\varphi$ 
 \be 
{\cal L} = \frac12 \varphi^T T {\cal O} \varphi 
\ee
which is invariant under a linear symmetry 
\be
\delta \varphi = K\varphi 
\ee
that generically acts also on the operator  $T {\cal O}$, which may depend on background fields.
The one-loop effective action can be regulated by subtracting a loop  of a massive
PV field $\phi$ with action 
\be 
{\cal L}_{\scriptscriptstyle PV} =  \frac12 \phi^T T  {\cal O} \phi +\frac12 M \phi^T T \phi 
\label{PV-l}
\ee 
where $M$ is a real parameter\footnote{To be precise, one should employ a set of PV fields with mass $M_i$ 
and relative weight $c_i$ in the loop to be able to regulate and cancel all possible one-loop divergences \cite{Pauli:1949zm}.
For simplicity, we consider only one PV field with relative weight $c=-1$, as this is enough  for our purposes.
The weight $c=-1$  means that we are subtracting a massive PV loop from the original one.}. 
The mass term identifies the operator $T$, that  in turn allows to find the operator ${\cal O}$.
As we shall see, in fermionic theories with a first order differential operator  ${\cal O}$ in the kinetic term, 
the operator ${\cal O}^2$ acts as a regulator in the final formula for the anomaly.
The invariance of the original action extends to an invariance of the massless part of the PV action by defining 
\be
\delta \phi = K \phi
\label{PV-t}
\ee
so that only the mass term may break the symmetry 
\be
\delta{\cal L}_{\scriptscriptstyle PV} =  \frac12 M \phi^T (T K +K^T T +\delta T) \phi =
M \phi^T (T K +\frac12 \delta T) \phi 
\;.
\ee

The path integral $Z$ and the one-loop effective action $\Gamma$ are regulated by the PV field
\be
Z=e^{i \Gamma} =\int D\varphi\; e^{i S}  \qquad \to \qquad Z=e^{i \Gamma} =
\int D\varphi D\phi\;  e^{i (S + S_{\scriptscriptstyle PV})} 
\ee
where it is understood that one should take the $M\to \infty$ limit, with all divergences canceled as explained in the footnote.
The anomalous response of the path integral under a symmetry 
is due to the PV mass term only, as one can define the measure of the PV field so to make the whole
path integral measure invariant \cite{Diaz:1989nx}. 
In a hypercondensed notation, where  a term like $\phi^T \phi$
includes in the sum of the (suppressed) indices a  spacetime integration as well, a lagrangian like the one in \eqref{PV-l}
is equivalent to the  action, and one may compute the symmetry variation of the regulated path integral to obtain
\begin{align}
i\delta \Gamma =i \la\delta S\ra  &=  \lim_{M \to \infty} \ i M \la \phi^T (TK +\frac12 \delta T) \phi \ra
\ccr
&= - \lim_{M \to \infty}  
{\rm Tr} \biggl [\biggl (K + \frac12 T^{-1} \delta T \biggr ) \biggl ( 
1+ \frac{\cal O}{M} \biggr )^{\!\!\! -1} \biggr ] 
\end{align}
where brackets $\la ...\ra$ denote normalized correlation functions.
For our purposes, it is convenient to cast it in an  equivalent form \cite{Hatsuda:1989qy}
 \be
i\delta \Gamma =i \la\delta S\ra  =
- \lim_{M \to \infty}  
{\rm Tr} \biggl [\biggl (K + \frac12 T^{-1} \delta T + \frac12 \frac{\delta {\cal O}}{M} \biggr ) 
\biggl ( 1- \frac{{\cal O}^2}{M^2} \biggr )^{\!\!\! -1} \biggr ] 
\label{3.8}
\ee
which is obtained 
by using the identity $ 1 = (1 - \frac{\cal O}{M}) (1 - \frac{\cal O}{M})^{-1}$  
and the invariance of  the massless action 
\be
\delta {\cal L} = \varphi^T \left ( T {\cal O} K +\frac12 \delta T {\cal O} +\frac12 T \delta  {\cal O} \right ) \varphi =0 \;.
\ee
In deriving  these expressions, we have considered a fermionic theory, used the PV propagator
\be
\la \phi \phi^T\ra = \frac{i}{T {\cal O} + T M} \;,
\ee
taken into account the opposite sign for the PV field in the loop, and considered an invertible mass matrix $T$. 
In the limit $M\to \infty$ the regulating term $ ( 1- \frac{{\cal O}^2}{M^2})^{-1}$ inside \eqref{3.8}
can be replaced by $e^{ \frac{{\cal O}^2}{M^2}}$. This is allowed  as for extracting the limit 
these regulators cut off the ultraviolet frequencies in an equivalent way 
(we assume that ${\cal O}^2$ is negative definite after a Wick rotation to euclidean space).
Clearly, if one finds a symmetrical mass term, then the symmetry would remain automatically anomaly free.

Heat kernel formulas may now be directly applied.  Denoting 
\be
J=K + \frac12 T^{-1} \delta T + \frac12 \frac{\delta {\cal O}}{M} \;, \qquad { R}=-{\cal O}^2
\label{jac}
\ee
the anomaly is related to the trace of the heat kernel  of the regulator  ${R}$ with an insertion of $J$
\be
i\delta \Gamma =i \la\delta S\ra  =
- \lim_{M \to \infty}   {\rm Tr} [J e^{ -\frac{R}{M^2}}] \;.
\label{tra}
\ee
This has the same form that appears in the original Fujikawa's method for computing 
anomalies \cite{Fujikawa:1979ay,Fujikawa:1980vr},
where $J$ is the infinitesimal part of the fermionic jacobian arising from a change of the path integral variables
under a symmetry transformation, and $R$ is the regulator. 
The limit extracts only the mass independent term (negative powers of the mass vanish in the limit, while positive 
(diverging) powers are made to cancel by using additional PV fields).
The PV method guarantees that the regulator $R$ together with $J$ produces consistent anomalies, which follows
from the fact that we are computing directly the variation of the effective action.
 
The heat kernel formulas that we need in the anomaly calculation are well-known, and we report them 
in appendix \ref{appB} using a minkowskian time. In particular, in four dimensions 
we just need the so-called Seeley-DeWitt coefficients $a_2({R})$ corresponding to the regulators $R$ associated to  
the different fields assembled into $\phi$. These are the only coefficients that survive in the limit ${M \to \infty}$
(as said, diverging pieces are removed by the PV renomalization).
Running through the various cases presented in the previous section, we can extract the ``jacobians" $J$  and regulators $R$ 
to find  the structure of the anomalies.
For the Weyl model we find
\begin{align}
& \partial_a \la J^a \ra =  \frac{ i }{(4\pi)^2}  \Big [\tr  [P_L a_{2}({R}_\lambda)]  -\tr [P_R a_{2} ({R}_{\lambda_c})]\Big ]
\ccr[2mm]
& 
\la T^a{}_a \ra =  -\frac{1}{2 (4\pi)^2} \Big [ \tr [P_L a_{2}({R}_\lambda)]  +  \tr[ P_R a_{2}({R}_{\lambda_c})]\Big ]  \;.
\label{W-anomaly}
\end{align}
These formulas are obtained by considering that for the $U(1)$ symmetry the jacobian $J$ in \eqref{jac} 
is extracted from the symmetry transformations of $\lambda$ and $\lambda_c$ in \eqref{axial-transf} 
\be
J= \left( \begin{array}{cc}   i\alpha P_L & 0  \\  0&  -i\alpha P_R      \end{array} \right)  \;.
\ee
Only $K$ contributes, as $\delta T$ vanishes  while we have neglected momentarily the contribution 
from $\delta {\cal O}$ (it vanishes after taking the traces in \eqref{W-anomaly}, as checked in the next section).
The infinitesimal parameter $\alpha$ is eventually factorized away from \eqref{tra} to obtain the local form in \eqref{W-anomaly}. 
In computing $J$ from \eqref{jac}, it is enough to check that the mass 
matrix $T$ is invertible on the relevant chiral spaces (extracted by the projectors $P_L$ and $P_R$).  
For the Weyl symmetry one uses instead the transformation laws  in \eqref{212} to find 
\be
J= \left( \begin{array}{cc}   \frac 12 \sigma P_L & 0  \\  0&  \frac 12 \sigma P_R      \end{array} \right)  \;,
\ee
where now it is crucial to consider that the covariant (under gravity) extension of the mass terms contains a factor of $e$, 
see eq. \eqref{M-mass-cov},
which brings in a contribution from  $\frac12 T^{-1} \delta T$ to $J$ ($\delta {\cal O}$  is neglected again for the same reason as before).
This contribution is necessary to guarantee that general coordinate invariance is kept anomaly free in the regularization.
The infinitesimal Weyl parameter $\sigma$ is then factorized away from \eqref{tra} to obtain 
the second equation in \eqref{W-anomaly}.

Proceeding in a similar way, we find for  the Dirac model  
\begin{align}
& \partial_a \la J^a \ra =  \frac{ i }{(4\pi)^2}  [\tr a_{2}({R}_\psi)  -\tr a_{2}({R}_{\psi_c})]
\ccr[2mm]
& \partial_a  \la J_5^a\ra =   
\frac{i}{(4\pi)^2} \Big [\tr [\gamma^5 a_{2} ({R}_\psi)] +\tr \gamma^5 [a_{2} ({R}_{\psi_c}) ] \Big]
\ccr[2mm]
& 
\la T^a{}_a \ra = -\frac{1}{2 (4\pi)^2}  [ \tr a_{2}({R}_\psi)  +\tr a_{2}({R}_{\psi_c}) ]  \;.
\label{D-anomaly}
\end{align}
All remaining traces are traces on the gamma matrices taken in the standard four dimensional Dirac spinor space.

\section{Anomalies}
\label{s4}

In this section  we compute systematically the chiral and trace anomalies for the Weyl and Dirac fermions described earlier.
We use, when applicable,  two different versions of the Pauli-Villars regularization with different mass terms. We verify that
the final results are consistent with each other, and coincide after taking into account the variation of local counterterms. 

\subsection{Chiral and trace anomalies of a Weyl fermion}

We consider first the case of a Weyl fermion. 

\subsubsection{PV regularization with Majorana mass}
\label{411}
 The regularization of the Weyl fermion coupled to an abelian gauge field is achieved in the most minimal way
 by using a PV fermion of the same chirality and with the Majorana mass term given in eq. \eqref{M-mass} added.
  This set-up was already used in \cite{Bastianelli:2016nuf}
  to address the case of a Weyl fermion in a gravitational background, but without the abelian gauge coupling. 
  The mass term is Lorentz invariant and does not introduce additional chiralities, but breaks the gauge and conformal 
  (and Weyl) symmetries. Therefore, one expects chiral and trace anomalies. 
  
To obtain the anomalies we have to compute the expressions in \eqref{W-anomaly} with the regulators contained 
inside the ${\cal O}^2$ given in eq. \eqref{PV-MM-reg}. They read
  \begin{align}
  {R}_\lambda &= - \DDslash(-A) \DDslash(A) P_L
  \ccr
 {R}_{\lambda_c} &= -\DDslash(A) \DDslash(-A) P_R \;.
  \end{align}
  Using the Seeley-DeWitt coefficients $a_2$ of these regulators,
 see appendix \ref{appC} for an outline of the calculation, we find for the chiral anomaly
\begin{align}
\partial_a \la J^a \ra =& 
\frac{1}{(4\pi)^2}\left ( \frac 16 \epsilon^{abcd}F_{ab}F_{cd} - \frac{8}{3}\partial_a(A^a A^2) + \frac 23 \Box(\partial A) \right ) 
\end{align}
where $F_{ab}=\partial_a A_b-\partial_b A_a$. It contains normal-parity terms that can be canceled by the gauge variation of the local counterterm
\be
\Gamma_{1} = \int \frac{d^4x}{(4\pi)^2} \, \left( \frac 23 A^4 - \frac{1}{3} A^a \Box A_a  \right ) 
\;,
\ee
so that the chiral gauge anomaly takes the form
\be
\partial_a \la J^a \ra = \frac{1}{96\pi^2} \epsilon^{abcd}F_{ab}F_{cd}  
\label{44}
\ee
which is the standard result.

Similarly, we compute the trace anomaly which is given by
\be
\la T^a{}_a  \ra = -\frac{1}{(4\pi)^2} \left (\frac 23 (\partial_a A_b)(\partial^a A^b) -\frac 23 (\partial A)^2  - \frac 23 \Box A^2 \right ) \;.
\ee
It does not contain any odd-parity contribution. Gauge invariance is broken by the chiral anomaly, still the trace anomaly 
can be cast in a gauge invariant form by varying a local counterterm with a Weyl  transformation 
and then restricting to flat space. The (gravity covariant and gauge noninvariant) counterterm is given by 
\be
\Gamma_{2} = \int \frac{d^4x \sqrt{g}}{(4\pi)^2}  \left(  \frac 13 (\nabla^\mu A^\nu)(\nabla_\mu A_\nu)  +\frac 16 R A^2 \right) 
\ee
and the trace anomaly takes the form
\be
\la T^a{}_a  \ra = -\frac{1}{48\pi^2}F_{ab}F^{ab} \;.
\ee

The counterterms $\Gamma_1$ and $\Gamma_2$ are consistent with each other, and merge into the 
unique counterterm (needed only at linear order in the metric) 
\be
\Gamma_{3} = \int \frac{d^4x \sqrt{g}}{(4\pi)^2}  \left(  \frac 23 A^4 + \frac 13 (\nabla^\mu A^\nu)(\nabla_\mu A_\nu)  +\frac 16 R A^2 \right) 
\ee 
where, of course, $A^2 = g^{\mu\nu} A_\mu A_\nu$ and $ A^4 =(A^2)^2$. 

  Thus, we have seen that the trace anomaly of a Weyl fermion does not contain any contribution from 
  the topological density $F\tilde F$ (which on the other hand enters the chiral anomaly in \eqref{44}, as well-known). 
Also, it can be presented in a gauge invariant form by the variation of a local counterterm, and 
equals half the standard trace anomaly of a Dirac fermion.
 These are the main results of our paper.

\subsubsection{PV regularization with Dirac mass}
\label{subsection}

For using a Dirac mass  we have to include also a right handed free fermion in the PV  lagrangian.
The lagrangian is given in \eqref{dirac-1}, and from eq. \eqref{PV-WDM-reg} one finds the regulators
  \begin{align}
  {R}_\lambda &= -   \dslash \DDslash (A)P_L
  \ccr
 {R}_{\lambda_c} &= -\dslash\DDslash (-A) P_R \;.
  \end{align}
   Then, from the corresponding heat kernel coefficients $a_2$ we find the chiral anomaly
 \be
\partial_a \la J^a \ra 
= \frac{1}{(4\pi)^2} \left (\frac{1}{6}\epsilon^{abcd}F_{ab}F_{cd} - \frac{1}{3}\partial_a (A^a A^2) + \frac{1}{3}\Box(\partial A)  \right ).
\ee
It contains noncovariant normal-parity terms, that are canceled by the variation of the local counterterm 
\be
\Gamma_{4} = \int \frac{d^4x}{(4\pi)^2} \, \left( \frac{1}{12}A^4 - \frac{1}{6} A^a \Box A_a 
\right) 
\ee
so that the anomaly takes the standard form
\be
\partial_a \la J^a \ra = \frac{1}{96\pi^2} \epsilon^{abcd}F_{ab}F_{cd} 
\ee
as in the previous section. 

Unfortunately, we cannot proceed to compute in a simple way the trace anomaly using this regularization, 
as the mass term breaks the Einstein and  local Lorentz symmetries as well.
The ensuing anomalies should then be computed and canceled by local counterterms, to find eventually the expected agreement 
of the remaining trace anomaly with the one found in the previous section. 

\subsection{Chiral and trace anomalies of a Dirac fermion}

We now consider the case of the massless Dirac spinor coupled to vector and axial gauge fields
with lagrangian given in eq. \eqref{D-lag}.
The most natural regularization is obtained by employing a Dirac mass for the PV fields,
 but  we also consider a Majorana mass. The latter allows to take a chiral limit in a simple way, 
 which we use to rederive the previous results on the Weyl fermion. 

\subsubsection{PV regularization with Dirac mass}

The relevant regulators are obtained from \eqref{PVD-Dmass}
and read
  \begin{align}
  {R}_\psi &= -    \DDslash(A,B)^2 
  \ccr
 {R}_{\psi_c} &= -  \DDslash(-A,B)^2  \;.
  \end{align} 
 The vector symmetry is guaranteed to remain anomaly free by the invariance of the mass term, 
 while the chiral anomaly from \eqref{D-anomaly}  becomes  
\begin{align}
\partial_a  \la J^a_5  \ra = \frac{1}{(4\pi)^2} &\left ( \epsilon^{abcd}F_{ab}(A)F_{cd}(A) +\frac{1}{3} \epsilon^{abcd}F_{ab}(B)F_{cd}(B) 
- \frac{16}{3} \partial_a ( B^a B^2) + \frac{4}{3}\Box (\partial B) \right )\;.
\end{align}
It contains normal-parity terms in the $B$ field. They are canceled by the variation 
of  a local counterterm 
\be
\Gamma_{5} = \int \frac{d^4x}{(4\pi)^2} \, \left(\frac{4}{3} B^4 - \frac{2}{3} B^a \Box B_a \right)
\ee
so that one ends up with
\begin{align}
&\partial_a  \la J^a  \ra = 0 \label{4Ddirac_vector_anomaly}\\
& \partial_a  \la J^a_5  \ra = \frac{1}{(4\pi)^2} \left( \epsilon^{abcd}F_{ab}(A)F_{cd}(A) +\frac{1}{3} \epsilon^{abcd}F_{ab}(B)F_{cd}(B) \right) \;.
\label{4Ddirac_chiral_anomaly}
\end{align}

As for the trace anomaly,  we find from \eqref{D-anomaly} 
		\begin{align}
\la T^a{}_a  \ra &= -\frac{1}{(4\pi)^2} \left( \frac 23 F_{ab}(A)F^{ab}(A) +\frac 43 (\partial_a B_b)(\partial^a B^b)  - \frac 43 (\partial B)^2  - \frac 43 \Box B^2 \right) 
\label{trace_dirac}\end{align}
and the counterterm
\be
\Gamma_{6} = \int \frac{d^4x  \sqrt{g}}{(4\pi)^2} \left(  \frac 23 (\nabla^\mu B^\nu)(\nabla_\mu B_\nu)  +\frac 13 R B^2 \right) 
\ee
brings it into the gauge invariant form
\begin{equation}
\la T^a{}_a  \ra =  -\frac{1}{24\pi^2} \Big (F_{ab}(A)F^{ab}(A) + F_{ab}(B)F^{ab}(B)  \Big ) \;.
\label{trace_dirac2}\end{equation}
All these counterterms merge naturally into the complete counterterm
\be
\Gamma_{7} = \int \frac{d^4x  \sqrt{g}}{(4\pi)^2} \left(  \frac{4}{3} B^4 +
\frac 23 (\nabla^\mu B^\nu)(\nabla_\mu B_\nu)  +\frac 13 R B^2 \right) 
\;.
\ee

\subsubsection{PV regularization with Majorana mass}

Finally, we consider the regularization with a Majorana mass.
As both vector and chiral symmetries are broken by the mass term, we expect anomalies in both $U(1)$ currents.
From eq. \eqref{PVD-MM} we find the regulators
  \begin{align}
  {R}_\psi &= -   \DDslash(-A,B)\DDslash(A,B) 
  \ccr
 {R}_{\psi_c} &= -  \DDslash(A,B)\DDslash(-A,B)  \;.
  \end{align}
 Thus,  we compute from  \eqref{D-anomaly}  
\begin{align}
\partial_a  \la J^a  \ra =\frac{1}{(4\pi)^2}&\left ( \frac 23  \epsilon^{abcd} F_{ab} (A) F_{cd}(B)
+\frac{4}{3}\Box(\partial A) 
-\frac{16}{3} \partial_a[A^a(A^2+B^2)]  -\frac{32}{3} \partial_a(B^a A_b B^b) \right )
\end{align}
and 
\begin{align}
\partial_a  \la J_5^a  \ra =\frac{1}{(4\pi)^2}&\left (\frac{1}{3} \epsilon^{abcd}F_{ab} (A) F_{cd}(A)
+ \frac{1}{3} \epsilon^{abcd}F_{ab} (B) F_{cd}(B)
+\frac{4}{3}\Box(\partial B)  \right.
\nonumber \\
&\left. -\frac{16}{3}  \partial_a[B^a(A^2+B^2)] -\frac{32}{3}  \partial_a(A^a A_b B^b)
\right ) \;.
\end{align}
The counterterm $\Gamma_{8}+\Gamma_{9}$
\begin{align}
\Gamma_{8}&=\int \frac{d^4x}{(4\pi)^2} \, \left (
\frac{4}{3} (A^2 +B^2)^2 +\frac{16}{3} (A^a B_a)^2 -\frac{2}{3} A^a\Box A_a -\frac{2}{3} B^a\Box B_a \right)
\ccr
\Gamma_{9} &= \int \frac{d^4x}{(4\pi)^2} \,  \left ( \frac{8}{3}\epsilon^{abcd} B_a A_b (\partial_c A_d)   \right)
\end{align}
 allows to recover vector gauge invariance, and the anomalies take the form
\begin{align}
&\partial_a  \la J^a  \ra = 0 \label{4Dmajorana_vector_anomaly}\\
& \partial_a  \la J^a_5  \ra = \frac{1}{(4\pi)^2} \left( \epsilon^{abcd}F_{ab}(A)F_{cd}(A) +\frac{1}{3} \epsilon^{abcd}F_{ab}(B)F_{cd}(B) \right) \;.
\label{4Dmajorana_chiral_anomaly}
\end{align}

As for the trace anomaly, we find
\begin{align}
\la T^a{}_a  \ra &= -\frac{1}{(4\pi)^2} \left (\frac 43 (\partial_a A_b)(\partial^a A^b) -\frac 43(\partial A)^2 - \frac 43 \Box A^2 +\frac 43 (\partial_a B_b)(\partial^a B^b)  - \frac 43 (\partial B)^2  - \frac 43 \Box B^2 \right )
\label{trace_majorana}\end{align}
and using the counterterm 
\be
\Gamma_{10} = \int \frac{d^4x  \sqrt{g} }{(4\pi)^2} \left(  
\frac 23 (\nabla^\mu A^\nu)(\nabla_\mu A_\nu)  + \frac 23 (\nabla^\mu B^\nu)(\nabla_\mu B_\nu) 
+\frac 13 R (A^2+B^2) \right) 
\ee
we get the final gauge invariant form
\begin{equation}
\la T^a{}_a  \ra =  -\frac{1}{24\pi^2} \Big ( F_{ab}(A)F^{ab}(A) + F_{ab}(B)F^{ab}(B)  \Big)\;.
\label{trace_majorana2}\end{equation}

The counterterms employed in this section are consistent with each other, and combine into a unique 
final counterterm, which we report for completeness
\begin{align}
\Gamma_{11} = \int \frac{d^4x  \sqrt{g} }{(4\pi)^2} &\left( 
\frac 23 (\nabla^\mu A^\nu)(\nabla_\mu A_\nu)  + \frac 23 (\nabla^\mu B^\nu)(\nabla_\mu B_\nu) 
+\frac 13 R (A^2+B^2) 
\right . \ccr
& \left . 
+ \frac{4}{3} (A^2 +B^2)^2 +\frac{16}{3} (A^\mu B_\mu)^2  + \frac{4}{3} \frac{\epsilon^{\mu\nu\rho\sigma}}{\sqrt{g}} 
B_\mu A_\nu F_{\rho\sigma}(A)
\right ) \;.
\end{align}

Evidently, the anomalies computed with the Majorana mass coincide with those obtained with the Dirac mass, 
after using local counterterms.

The results of this section can be projected consistently to recover the chiral and trace anomalies  of the Weyl fermion.
Indeed, one can consider the limit $A_a = B_a \to  \frac 12 A_a$.
In this limit, a chiral projector $P_L = \frac{1+\gamma^5}{2} $ emerges
inside the Dirac lagrangian \eqref{D-lag} to reproduce the Weyl lagrangian \eqref{Weyl-act}.
In addition, in the coupling to gravity, the right handed  component of the Dirac field can be kept free
both in the kinetic and in the PV mass term, while preserving the covariance of the mass term for the left handed 
part of the PV Dirac fermion. Thus, the right handed part can be ignored altogether.
Indeed, one may verify that the anomalies in subsection 
\ref{411} are reproduced by those computed here, including the counterterms,
by setting $A_a = B_a \to  \frac 12 A_a$  (note that the current $J^a$ in \ref{411} 
corresponds to half the sum of $J^a$ and $J^a_5$ of this section).

Finally, we have checked that terms proportional to $\delta \cal{O}$ in \eqref{jac}
never contribute to the anomalies computed thus far, as  the extra terms vanish under the Dirac trace.

\section{Conclusions}
\label{s5}
We have calculated the trace anomaly of a Weyl fermion coupled to an abelian gauge field. 
We have found that the anomaly does not contain any odd-parity contribution. In particular, we have shown 
that the Chern-Pontryagin term $F\tilde F$ is absent,
notwithstanding the fact that it satisfies the consistency conditions for Weyl anomalies.
The chiral anomaly implies that gauge invariance is broken. Nevertheless the trace anomaly can be cast in a gauge invariant form, equal to half the standard contribution of a nonchiral Dirac fermion.  

While this result seems to have no direct implications for the analogous case in curved background, 
it strengthens the findings of 
ref. \cite{Bastianelli:2016nuf}\footnote{A confirmation of those results has also appeared recently in \cite{Frob:2019dgf}.}. 

Recently, a generalized axial metric background has been developed in \cite{Bonora:2017gzz, Bonora:2018obr}
to motivate and explain the appearance of the Pontryagin term in the trace anomaly of a Weyl fermion, which however
is in contradiction with the explicit calculation presented in \cite{Bastianelli:2016nuf}. Perhaps it would be useful to apply 
the methods used here in the context of the axial metric background to clarify the situation, 
and spot the source of disagreement.

\acknowledgments{We thank Stefan Theisen for extensive discussions, suggestions, critical reading of the manuscript,
 and hospitality at AEI. In addition, FB would like to thank Loriano Bonora and Claudio Corian\`o for useful discussions, 
and MB acknowledges Lorenzo Casarin for interesting discussions and insights into anomalies and heat kernel.}

\appendix 

\section{Conventions}
\label{appA} 

We use a mostly plus Minkowski metric $\eta_{ab}$.  
The Dirac matrices $\gamma^a$ satisfy 
\be
\{ \gamma^a, \gamma^b \} =2 \eta^{ab} 
\ee
and the conjugate Dirac spinor $\bpsi$ is defined using $\beta= i\gamma^0$ by
\be 
\bpsi = \psi^\dagger \beta \;.
\ee
The hermitian chiral matrix $\gamma^5$ is given by
\be 
\gamma^5=- i \gamma^0 \gamma^1\gamma^2 \gamma^3
\ee
and used to define the chiral projectors
\be
P_L = \frac{\mathbb{1}+\gamma_5}{2} \ , \qquad P_R = \frac{\mathbb{1}-\gamma_5}{2} 
\ee
that split a Dirac spinor $\psi$ into its left and right  Weyl components
\be
\psi = \lambda +\rho \;, \qquad \lambda=P_L \psi \;,  \quad \rho=P_R \psi  \;.
\ee
The charge conjugation matrix $C$ satisfies
\be
C\gamma^{a} C^{-1} = -\gamma^{a T }  \;,
\label{C-matrix}
\ee
it is antisymmetric and used to define the charge conjugation of  the spinor $\psi$ by
\be
\psi_c = C^{-1} \bpsi^T 
\ee
for which the roles of particle and antiparticle get interchanged.
Note that a chiral spinor  $\lambda$ has its charge conjugated field $\lambda_c$  of opposite chirality.
A Majorana spinor $\mu$ is a spinor  that equals its charged conjugated spinor
\be
\mu =\mu_c \;.
\label{Majc}
\ee
This constraint is incompatible with the chiral constraint, and Majorana-Weyl  spinors do not exist in 4 dimensions.

We find it convenient, as a check on our formulas, to use the  chiral representation of the gamma matrices.
In terms of $2\times 2$  blocks they are given by
\be
\gamma^0= -i
\left( \begin{array}{cc}
 0 & \mathbb{1} \\ 
\mathbb{1} & 0
\end{array} \right) \ , \qquad 
\gamma^i = -i  \left( \begin{array}{cc}
0 &  \sigma^i \\ 
- \sigma^i & 0
\end{array} \right) 
\ee
where $\sigma^i$ are the Pauli matrices, so that 
 \be
\gamma^5=
\left( \begin{array}{cc}
\mathbb{1} & 0 \\ 
0 & - \mathbb{1}
\end{array} \right) 
\;, \qquad
\beta = i\gamma^0 = 
\left( \begin{array}{cc}
0 & \mathbb{1} \\ 
\mathbb{1} & 0
\end{array} \right) \;.
\ee
The chiral representation makes evident that the Lorentz generators in the spinor space 
$M^{ab}=\frac{1}{4} [\gamma^a,\gamma^b] = \frac12 \gamma^{ab}  $
take a block diagonal form
\be
M^{0i} = \frac12 
\left( \begin{array}{cc}
\sigma^i & 0\\ 
0&-\sigma^i  
\end{array} \right) \;, \qquad 
M^{ij} = \frac{i}{2} \epsilon^{ijk} 
\left( \begin{array}{cc}
\sigma^k & 0 \\ 
 0 & \sigma^k
\end{array} \right)  
\ee
and do not mix the chiral components of  a Dirac spinor (as $\gamma^5$ is also block diagonal). 
The usual two-dimensional Weyl spinors appear inside a four-dimensional Dirac spinor as follows 
\be
\psi = 
\left( \begin{array}{c}
l \\ 
r 
\end{array} \right)  \;, \qquad
\lambda = 
\left( \begin{array}{c}
l \\ 
0
\end{array} \right)  \;, \qquad
\rho = 
\left( \begin{array}{c}
0 \\ 
r 
\end{array} \right)  
\label{A.12}
\ee
where  $l$ and $r$ indicate  two-dimensional independent spinors of opposite chirality.
In the chiral representation one may take the charge conjugation matrix $C$ to be given by
\be
C = \gamma^2 \beta = -i 
\left( \begin{array}{cc}
\sigma^2 & 0\\ 
0&-\sigma^2  
\end{array} \right)  
\ee
and satisfies 
\be
C=-C^T=-C^{-1}=-C^\dagger =C^* 
\ee
(some of these relations are  representation dependent).
In the chiral representation the Majorana constraint \eqref{Majc} takes the form 
\be
\mu=\mu_c    \qquad \to  \qquad  
\left( \begin{array}{c} l \\ r \end{array} \right)  =
\left( \begin{array}{c} i \sigma ^2 r^* \\ - i \sigma ^2 l^* \end{array} \right)  
\ee
which shows that the two-dimensional spinors $l$ and $r$ cannot be independent. 
The Majorana condition can be solved in terms of the single two-dimensional left-handed spinor $l$ as 
\be
\mu = \left( \begin{array}{c} l \\  -i \sigma^2 l^* \end{array} \right) 
\ee
which, evidently, contains the four-dimensional chiral spinors $\lambda$ and $\lambda_c$ defined by
\be
\lambda = \left( \begin{array}{c} l \\  0 \end{array} \right) 
\;, \qquad
\lambda_c = \left( \begin{array}{c} 0 \\  -i \sigma^2 l^* \end{array} \right)  \;.
\ee
In a four-dimensional spinors notation one can write
\be
\mu =\lambda +\lambda_c \;.
\ee
Alternatively, the Majorana condition can be solved in terms of the two-dimensional right-handed spinor $r$ as
\be
\mu = \left( \begin{array}{c} i \sigma^2 r^* \\  r \end{array} \right ) 
\ee
which contains the four-dimensional chiral spinors $\rho$ and $\rho_c$
\be
\rho = \left( \begin{array}{c} 0  \\    r  \end{array} \right) 
\;, \qquad
\rho_c = \left( \begin{array}{c}  i \sigma^2 r^*  \\  0 \end{array} \right) 
\ee
and $\mu = \rho+\rho_c$.  This solution is of course the same as the previous one, 
as one may identify $\lambda=\rho_c$.

The explicit dictionary between Weyl and Majorana spinors shows clearly that the field theory 
of a Weyl spinor is equivalent to that of a Majorana spinor, as Lorentz symmetry fixes uniquely their actions, which are bound to be identical.

Finally, we normalize our $\epsilon$ symbols by $\epsilon_{0123}=-1$ and $\epsilon^{0123}=1$, so that
\be
\frac14
\tr(\gamma^5 \gamma^a\gamma^b\gamma^c\gamma^d) = i \epsilon^{abcd} \;.
\ee
 
\section{The heat kernel}
\label{appB} 

We consider an operator in flat $D$ dimensional spacetime of the form  
\be
H= -\nabla^2 + V
\label{B1}
\ee
with $V$ a matrix potential and $\nabla^2=\nabla^a \nabla_a$ constructed with a 
gauge covariant derivative  $\nabla_a=\partial_a + W_a$ that satisfies
\be
[\nabla_a,\nabla_b] = \partial_a W_b-\partial_b W_a + [W_a, W_b] ={\cal F}_{ab}  
\label{B2}
\;.
\ee

The trace of the corresponding heat kernel is perturbatively given by
\begin{align}
{\rm Tr} \left [ J  e^{-i s H} \right ] 
&=
\int d^Dx \, \tr  \left  [J(x) \la x|e^{-i s H} | x \ra \right] 
\\
&= \int d^Dx \,
\frac{i}{(4 \pi i s)^{\frac{D}{2}}} \sum_{n=0}^\infty \tr [ J(x)a_n(x,H)] (is)^n
\ccr
& =\int d^Dx\, 
\frac{i}{(4 \pi i s)^\frac{D}{2}}\, {\rm tr}\, [J(x)  (a_0(x,H)+ a_1(x,H) i s + a_2(x,H) (is)^2+...)] 
\nonumber
\end{align}
where the symbol ``tr'' is the trace on the remaining discrete matrix indices, 
$J(x)$ is an arbitrary matrix function,
and $a_n(x,H)$ are the so-called Seeley-DeWitt coefficients, or heat kernel coefficients.  
They are matrix valued, and the first few ones are given by
\bea
a_0(x,H) \eqa {\mathbb 1} \ccr
a_1(x,H) \eqa    - V \ccr
a_2(x,H) \eqa  \frac12 V^2  -\frac{1}{6} \nabla^2 V + \frac{1}{12} {\cal F}_{ab}^2 \,.  \qquad
\label{SdW}
\eea
As $V$ is allowed to be a matrix, then $\nabla_a V= \partial_a V+ [W_a, V]$, etc..

In the main text, the role of the hamiltonian $H$ is played by the various regulators $R$,
and $ is \sim \frac{1}{M^2}$, see eq. \eqref{tra}. In $D=4$ the $s$ independent term is precisely the one 
with $a_2(x,H)$, which is the coefficient producing the anomalies in 4  dimensions 
(we use a minkowskian  set-up,  but justify the heat kernel formulas by Wick rotating to an euclidean time and back, when necessary).

More details on the heat kernel expansion are found in  \cite{DeWitt:1965jb, DeWitt:1985bc}, where the 
coefficients appear with the additional coupling to a background metric. 
They have been recomputed with quantum  mechanical path integrals in \cite{Bastianelli:1992ct},
a useful report is \cite{Vassilevich:2003xt}, while in \cite{Bastianelli:2000hi} one may find the 
explicit expression for $a_3(x,H)$, originally calculated by Gilkey \cite{Gilkey:1975iq}, which
is relevant for calculations of anomalies in 6 dimensions.

\section{Sample calculations}
\label{appC} 

As an example of the calculations leading to the results  of section  \ref{s4}, we consider the case
of the PV regularization with Majorana mass used for the Weyl model in section \ref{411}.
One regulator needed there  is 
\be
  {R}_\lambda = - \DDslash(-A) \DDslash(A) P_L \;.
    \ee
Neglecting the projector, that can be reinstated later, one should cast it in the general form of eq.  \eqref{B1}.
Expanding the covariant derivatives in the latter one finds
\be
H = - \nabla ^2 +V = -\partial^a\partial_a - 2W^a\partial_a -(\partial_a W^a) - W^aW_a +V \;.
\label{C2}
\ee
Similarly, by expanding ${R}_\lambda$ one finds (up to the projector)
\begin{align}
{R}_\lambda &=  - \DDslash(-A) \DDslash(A) = -D^a(-A)  D_a(A)  -2i \gamma^{ab} A_a\partial_b +\frac{i}{2} F_{ab} \gamma^{ab}
\ccr
&= -\partial^a\partial_a +2i\gamma^{ab}A_b\partial_a  + i(\partial^a A_a) - A^aA_a + \frac{i}{2} F_{ab} \gamma^{ab}
\label{C3}
  \end{align}
  where $\gamma^{ab} =\frac12 [\gamma^a,\gamma^b]$.
Comparing \eqref{C2} and \eqref{C3} one fixes
\begin{align}
W^a &= -i\gamma^{ab}A_b
\ccr
V &= 2 A^a A_a +i(\partial^a A_a)  \;.
  \end{align}
At this stage one proceeds to evaluate the field strength ${\cal F}_{ab}$ in eq. \eqref{B2}
associated to this particular $W^a$, and use it to evaluate the 
coefficient $a_2({R}_\lambda)$ from $a_2(H)$ of eq. \eqref{SdW} (remembering to reinsert the projector).
In particular, evaluating the trace  one finds 
\begin{align}
\tr  [P_L a_{2}({R}_\lambda)]  &= 
\frac 23 (\partial_a A_b)(\partial^a A^b) -\frac 23 (\partial A)^2  - \frac 23 \Box A^2 \ccr
&+i 
 \left (  \frac{4}{3}\partial_a(A^a A^2) - \frac 13 \Box(\partial A) -\frac{1}{12} \epsilon^{abcd}F_{ab}F_{cd}
\right )  \;.
\end{align}
 Note that this particular coefficient contains an odd-parity term proportional
to the topological density $F\tilde F$.
Similarly, one computes the coefficients related the other regulators, and proceeds to evaluate  \eqref{W-anomaly}
and \eqref{D-anomaly}.

We have checked our trace calculations on the gamma matrices also by computer, employing a notebook 
developed in~\cite{Clifford_Algebra_Traces} using the xAct and xTensor packages~\cite{xAct,xTensor}.


\end{document}